\documentclass[11pt]{article}

\usepackage[a4paper,margin=1.9cm]{geometry}
\usepackage{amsmath,amssymb,enumerate}
\usepackage{hyperref}
\hypersetup{colorlinks=true,breaklinks=true,allcolors=blue}
\usepackage{float}
\usepackage{authblk}
\usepackage[font=small]{caption}
\usepackage{graphicx}
\newcommand\revision[1]{#1}

\title{\revision{Mean exit time in irregularly-shaped annular\\ and composite disc domains}}
\author[1]{Elliot~J. Carr\footnote{To whom correspondence should be addressed. E-mail: \href{elliot.carr@qut.edu.au}{elliot.carr@qut.edu.au}}}
\author[1]{Daniel~J. VandenHeuvel}
\author[1]{Joshua~M. Wilson}
\author[1]{Matthew~J. Simpson}
\affil[1]{School of Mathematical Sciences, Queensland University of Technology, Brisbane, Queensland 4001, Australia}

\date{}

\renewcommand{\epsilon}{\varepsilon}

\newcommand{\matlab}{MATLAB}
\begin{document}

\maketitle
\bigskip\medskip
\begin{abstract}
Calculating the mean exit time (MET) for models of diffusion is a classical problem in statistical physics, with various applications in biophysics, economics and heat and mass transfer. While many exact results for MET are known for diffusion in simple geometries involving homogeneous materials, calculating MET for diffusion in realistic geometries involving heterogeneous materials is typically limited to repeated stochastic simulations or numerical solutions of the associated boundary value problem (BVP). In this work we derive exact solutions for the MET in irregular annular domains, including some applications where diffusion occurs in heterogenous media. These solutions are obtained by taking the exact results for MET in an annulus, and then constructing various perturbation solutions to account for the irregular geometries involved.  These solutions, with a range of boundary conditions, are implemented symbolically and compare very well with averaged data from repeated stochastic simulations and with numerical solutions of the associated BVP.  Software to implement the exact solutions is available on \href{https://github.com/ProfMJSimpson/Exit_time}{GitHub}.
\end{abstract}


\section{Introduction} \label{intro}
A fundamental property of diffusive transport phenomena is the concept of mean exit time (MET), which is an application of the more general concept of the first passage time~\cite{Redner2001,Krapivsky2010,Hughes1995}.  Estimates of MET provide insight into the timescale required for a diffusing particle to reach a certain target, such as an absorbing boundary.  From a macroscopic point of view, the MET is related the mean action time which gives a simple, finite approximation for the duration of a diffusive process~\cite{Ellery2012a,Ellery2012b}.  Exact expressions for MET are well known for diffusion in simple geometries, such as lines, discs and spheres~\cite{Redner2001,Krapivsky2010}.  Generalising to more complicated geometries, such as wedges~\cite{Di2008,Chupeau2015}, symmetric domains~\cite{Vaccario2015,Rupprecht2015,Carr2020,Godec2015}, elongating domains~\cite{Simpson2015a,Simpson2015b}, slender domains~\cite{Kurella2014,Lindsay2015,Grebenkov2019}, \revision{simply connected irregular domains \cite{Grebenkov2016}}, small targets~\cite{Lindsay2017,Grebenkov2020}, and certain features of microscopic transport processes~\cite{Lotstedt2015,Meinecke2016,Wardak2020,Padash2020} are active areas of research.  Without such exact results, we must rely on computational insights obtained by performing a large number of stochastic simulations or by numerically solving the appropriate continuum-level partial differential equation (PDE) model.  Neither of these approaches provide the general mathematical insight offered by exact solutions.

\revision{Our previous work in this area \cite{Simpson2021} focused} on deriving exact solutions for the MET on a range of irregular two--dimensional (2D) geometries by using classical results for the MET on a disc and ellipse, and then constructing perturbation solutions~\cite{Farlow82,McCollum2014}.  Such approaches allow us to approximate an irregular 2D domain as a perturbation of a disc or ellipse, and then solve the underlying boundary value problem (BVP) for the MET by projecting the boundary conditions for the irregular domain onto the unperturbed boundary~\cite{Simpson2021}.  This approach avoids the need for numerical calculations since the perturbation solutions can be evaluated symbolically.  This previous work, however, was limited to relatively simple problems involving diffusion in homogeneous media involving a single absorbing boundary. Here we provide further generalisations by considering more realistic scenarios that include absorbing and reflecting boundaries and diffusion in heterogeneous media, leading to BVPs with more complicated boundary conditions.

\begin{figure}[H]
	\centering
	\includegraphics[width=0.89\textwidth]{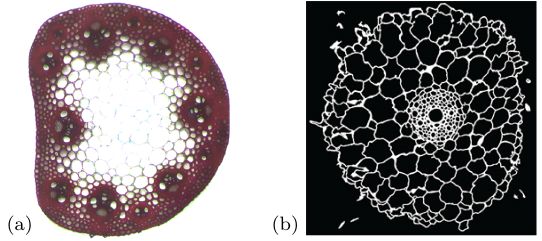}
	\caption{\textbf{Naturally-occurring heterogeneous annular domains.} The cross section of the plant stem in (a) takes the form of a perturbed annulus, where both the inner and outer boundary are irregular.  The cross section of the plant root in (b) takes the form of a compound disc, where the interface separating the different tissue types is irregular.  Images in (a) and (b) are reproduced from \cite{Matos2013} and \cite{Chopin2015}, respectively.}
	\label{fig1}
\end{figure}

To motivate the present study, we note that many biological tissues, within which chemical species undergo diffusive transport, involve irregular shapes, such as the cross sections of plant stems and plant roots in Figure \ref{fig1}.  The cross section of the plant stem in Figure \ref{fig1}(a) is annular in shape, with irregular inner and outer boundaries, whereas the cross section of the plant root in Figure \ref{fig1}(b) is shaped like a disc with a layered, irregular internal structure, where each layer is composed of different tissue types.  In this work we derive exact solutions for the MET in these kinds of irregular, heterogeneous domains.  First, starting with the known exact solution for the MET on an annulus, we derive new solutions for the MET on an annular region that is constructed by perturbing both the inner and outer boundaries of an annulus. In this case we consider different combinations of boundary conditions on the inner and outer boundaries.  Second, starting with the known exact solution for the MET on a heterogeneous compound disc, we derive solutions for the MET in an irregular heterogeneous domain that is obtained by perturbing the location of the interface in a compound disc. We solve the resulting BVPs and provide solutions to an arbitrary number of terms in the perturbation solutions.  Evaluating these solutions symbolically, we show that the perturbation solutions compare very well with both numerical solutions of the governing BVP and with averaged data from repeated stochastic simulations.  MATLAB code available at \href{https://github.com/ProfMJSimpson/Exit\_time}{https://github.com/ProfMJSimpson/Exit\_time} is provided implementing (i) the perturbation solutions using symbolic computation; (ii) the numerical finite volume solution of the BVP; and, (iii) the stochastic random walk algorithm.

\medskip

\section{Results and Discussion} \label{Results}
Standard arguments for unbiased random walk models on domains with absorbing boundary conditions can be used to show that the MET is given by the solution of a linear elliptic partial differential equation~\cite{Redner2001,Krapivsky2010,Hughes1995}:
\begin{align}
\label{eq:GovEq}
\nabla \cdot (D\nabla T(\mathbf{x})) &= -1, \quad \mathbf{x}\in\Omega,
\end{align}
where $D>0$ is the diffusivity and $T(\mathbf{x})$ is the MET for a particle released at location $\mathbf{x} = (x,y)\in\Omega$. Here, the diffusivity is related to the random walk model, $D = P \delta^2 / (4 \tau)$, where $\delta>0$ is the step length, $\tau>0$ is the duration between steps and $P \in (0,1]$ is the probability that a particle at location $\mathbf{x}$ attempts to undergo a step of length $\delta$ in the time duration $\tau$~\cite{Redner2001,Krapivsky2010,Hughes1995}.  The continuum description is valid in the constrained limit  $\delta \to 0$ and $\tau \to 0$, with $\delta^2/\tau$ held constant~\cite{Redner2001,Krapivsky2010,Hughes1995}.

In this work we seek solutions of Equation (\ref{eq:GovEq}) for various choices of $\Omega$ subject to absorbing or reflecting conditions imposed on the boundary of $\Omega$. In sections \ref{sec:annulus} and \ref{sec:compound_disc} we review known analytical solutions to Equation (\ref{eq:GovEq}) when $\Omega$ is an annulus and compound disc, respectively. In sections \ref{sec:perturbed_annulus} and \ref{sec:perturbed_compound_disc} we develop new approximate analytical solutions of Equation (\ref{eq:GovEq}) when $\Omega$ is an annulus with perturbed boundaries and a compound disc with a perturbed interface, respectively. \revision{For the compound disc, which does not have an inner boundary, we assume an absorbing outer boundary while for the annulus we consider three cases (i) absorbing inner and outer boundaries (ii) reflecting inner boundary and absorbing outer boundary and (iii) absorbing inner boundary and reflecting outer boundary.} Due to the nature of these domains, we work in polar coordinates when developing analytical solutions, writing $T(\mathbf{x})\equiv T(r,\theta)$, with position defined by the radial coordinate $r$ and angular coordinate $\theta$. \revision{In all cases we assume each perturbed boundary/interface can be described by a single value function of $\theta$, that is, any ray drawn from the origin intersects precisely one point of the boundary/interface.} The veracity of all analytical solutions are verified using appropriately averaged data from stochastic random walk simulations as well as a numerical solution of Equation (\ref{eq:GovEq}), discussed in Appendices \ref{app:finite_volume_solution}--\ref{app:stochastic_model}.

\subsection{Annulus} \label{sec:annulus}
We first review the case of an annulus where $\Omega = \{R_{1}<r<R_{2}\}$ and $P$ (and hence $D$) is constant across $\Omega$. When the inner and outer boundary at $r = R_{1}$ and $r = R_{2}$ are entirely absorbing or entirely reflecting, the MET depends on $r$ only and satisfies the differential equation~\cite{Carr2020}
\begin{gather}
\label{eq:a_ode}
\frac{D}{r}\frac{\text{d}}{\text{d}r}\left(r\frac{\text{d}T}{\text{d}r}\right) = -1, \quad  R_{1} < r < R_{2}.
\end{gather}
The solution, $T(r)$, of Equation (\ref{eq:a_ode}) admits a simple closed-form expression depending on the boundary conditions imposed:
\begin{enumerate}[(i)]
\item absorbing inner and outer boundaries ($T(R_{1}) = 0$ and $T(R_{2}) = 0$)
\begin{gather}
\label{eq:a_sol1}
T(r) = \frac{(R_{2}^{2}-R_{1}^2)\log(r) +R_{1}^{2}\log(R_{2})-R_{2}^{2}\log(R_{1})}{4D\log(R_{2}/R_{1})} - \frac{r^2}{4D},
\end{gather}
\item reflecting inner boundary and absorbing outer boundary ($\mathrm{d}T(R_{1})/\mathrm{d}r = 0$ and $T(R_{2}) = 0$)
\begin{gather}
\label{eq:a_sol2}
T(r) = \frac{R_{2}^{2}-r^{2}}{4D} + \frac{R_{1}^{2}}{2D}\log(r/R_{2}),
\end{gather}
\item \revision{absorbing inner boundary and reflecting outer boundary ($T(R_{1}) = 0$ and $\mathrm{d}T(R_{2})/\mathrm{d}r = 0$)}
\begin{gather}
\label{eq:a_sol3}
\revision{T(r) = \frac{R_{1}^{2}-r^{2}}{4D} + \frac{R_{2}^{2}}{2D}\log(r/R_{1})}.
\end{gather}
\end{enumerate}

\subsection{Perturbed annulus}
\label{sec:perturbed_annulus}
We now consider the case of a perturbed annulus where $\Omega = \{\mathcal{R}_{1}(\theta) < r < \mathcal{R}_{2}(\theta)\}$ and $P$ (and hence $D$) is constant across $\Omega$. Here, the inner and outer boundaries are described by polar curves, $r = \mathcal{R}_{1}(\theta)$ and $r = \mathcal{R}_{2}(\theta)$, where $\mathcal{R}_{1}(\theta) = R_{1}(1 + \epsilon g_{1}(\theta))$, $\mathcal{R}_{2}(\theta) = R_{2}(1 + \epsilon g_{2}(\theta))$, $\epsilon \ll 1$ is the perturbation parameter, $R_{1}$ is the unperturbed inner radius, $R_{2}$ is the unperturbed outer radius, and $g_{1}(\theta)$ and $g_{2}(\theta)$ are smooth $\mathcal{O}(1)$ periodic functions with period $2\pi$. For this problem, the solution of Equation (\ref{eq:GovEq}) can be expanded in powers of $\varepsilon$:
\begin{gather}
\label{eq:ann_T_expansion}
T(r, \theta) = \sum_{k=0}^\infty \epsilon^k T_k(r, \theta),
\end{gather}
where $T_0(r, \theta)$ satisfies Equation (\ref{eq:GovEq}) on the unperturbed annulus domain and $T_{k}(r,\theta)$ ($k = 1,2,3,\hdots$) satisfy Laplace's equation on the unperturbed annulus domain \cite{Simpson2021}. In the following sections, we derive exact expressions for $T_k(r, \theta)$ for all $k = 0,1,2,\hdots$, considering \revision{three} cases: (i) both absorbing inner and outer boundaries (ii) a reflecting inner boundary and an absorbing outer boundary \revision{and (iii) an absorbing inner boundary and a reflecting outer boundary.}

\subsubsection{Absorbing inner and outer boundaries}
\label{sec:perturbed_annulus1}
We first consider the case of absorbing inner and outer boundaries with $T = 0$ on $r = \mathcal{R}_{1}(\theta)$ and $r = \mathcal{R}_{2}(\theta)$:
\begin{gather}
T(R_{1} + R_{1} \epsilon g_{1}(\theta), \theta) = 0, \\
T(R_{2} + R_{2} \epsilon g_{2}(\theta), \theta) = 0.
\end{gather}
Expanding these equations in Taylor series centered at $r = R_{1}$ and $r = R_{2}$, respectively, gives
\begin{align*}
\sum_{i=0}^\infty \frac{(\epsilon R_{1} g_{1}(\theta))^i}{i!}\dfrac{\partial^i T}{\partial r^i}\left(R_{1},\theta \right) = 0, \\
\sum_{i=0}^\infty \frac{(\epsilon R_{2} g_{2}(\theta))^i}{i!}\dfrac{\partial^i T}{\partial r^i}\left(R_{2},\theta \right) = 0.
\end{align*}
Introducing the perturbation expansion (\ref{eq:ann_T_expansion}) and equating powers of $\varepsilon$ yields the following boundary conditions for each term in the perturbation solution (\ref{eq:ann_T_expansion}):
\begin{alignat}{2}
\mathcal{O}(1):\quad T_0(R_{1},\theta) &= 0, &\quad& T_0(R_{2},\theta) = 0,\\
\mathcal{O}(\varepsilon^{k}):\quad T_k(R_{1}, \theta) &= z_{k}(\theta), &\quad& T_k(R_{2}, \theta) = w_{k}(\theta),\quad k=1,2,3,\ldots,
\end{alignat}
where we have defined
\begin{gather*}
z_{k}(\theta) = -\sum_{i=1}^k \frac{(R_{1}g_{1}(\theta))^i}{i!}\dfrac{\partial^{i} T_{k-i}}{\partial r^i}\left(R_{1},\theta \right)\quad\text{and}\quad 
w_{k}(\theta) = -\sum_{i=1}^k \frac{(R_{2}g_{2}(\theta))^i}{i!}\dfrac{\partial^{i} T_{k-i}}{\partial r^i}\left(R_{2},\theta \right).
\end{gather*}

In summary, we have derived a BVP for each term in the perturbation solution (\ref{eq:ann_T_expansion}). The first term $T_{0}(r,\theta)\equiv T_{0}(r)$ depends upon the radial coordinate $r$ only and satisfies:
\begin{gather*}
\frac{D}{r}\frac{\text{d}}{\text{d}r}\left(r\frac{\text{d}T_0}{\text{d}r}\right) = -1,  \quad  R_{1}<r<R_{2},\\
 T_0(R_{1}) =0,\quad T_0(R_{2}) = 0,
\end{gather*}
which has solution
\begin{gather}
\label{eq:ann1_T0}
T_{0}(r) = \frac{(R_{2}^{2}-R_{1}^2)\log(r) +R_{1}^{2}\log(R_{2})-R_{2}^{2}\log(R_{1})}{4D\log(R_{2}/R_{1})} - \frac{r^2}{4D}.
\end{gather}
In contrast, the higher-order terms $T_{k}(r,\theta)$ ($k = 1,2,3,\hdots)$ in the perturbation expansion (\ref{eq:ann_T_expansion}) depend on both $r$ and $\theta$ and satisfy
\begin{gather}
\label{eq:ann1_Tk_pde}
\nabla^2 T_{k}= 0, \quad  R_{1}<r<R_{2},\\
\label{eq:ann1_Tk_bcs}
T_k(R_{1}, \theta) = z_{k}(\theta),\quad
T_k(R_{2}, \theta) = w_{k}(\theta).
\end{gather}
\revision{Note that the leading order term (\ref{eq:ann1_T0}) is identical to the corresponding solution on the unperturbed domain (\ref{eq:a_sol1}). This will always be the case for the perturbation solution (as we will see repeatedly throughout this work) so the MET on the unperturbed domain can always be recovered from the perturbation solution by simply setting $\varepsilon=0$. It follows from this observation that the higher-order terms can be interpreted as correction terms accounting for the projection of the boundary conditions from the perturbed annulus boundaries $r = \mathcal{R}_{1}(\theta)$ and $r = \mathcal{R}_{2}(\theta)$ onto the unperturbed annulus boundaries $r=R_{1}$ and $r=R_{2}$.}

Solving the BVP (\ref{eq:ann1_Tk_pde})--(\ref{eq:ann1_Tk_bcs}) using separation of variables gives:
\begin{gather}
\label{eq:ann1_T_fourier}
T_k(r, \theta) = c_0 + d_0 \log r + \sum_{n=1}^\infty \left[\left(c_n r^n + d_n r^{-n}\right) \cos(n\theta) + \left(a_n r^n + b_n r^{-n}\right) \sin(n\theta)\right],
\end{gather}
for $k=1,2,3,\ldots,$ where
\begin{gather*}
c_0  =  \dfrac{1}{2\pi \log(R_{2}/R_{1})}\int_0^{2\pi} z_{k}(\theta) \log(R_{2}) - w_{k}(\theta) \log(R_{1}) \, \textrm{d}\theta, \\
d_0  =  \dfrac{1}{2\pi \log(R_{2}/R_{1})}\int_0^{2\pi} w_{k}(\theta) -z_{k}(\theta) \, \textrm{d}\theta, \\
a_n  =  \dfrac{1}{\pi(R_{1}^{ 2n}-R_{2}^{ 2n})}\int_0^{2\pi} \left[R_{1}^{n}z_{k}(\theta)-R_{2}^{n}w_{k}(\theta) \right] \sin(n\theta) \, \textrm{d}\theta, \\
b_n  =  \dfrac{1}{\pi(R_{1}^{-2n}-R_{2}^{-2n})}\int_0^{2\pi} \left[R_{1}^{-n}z_{k}(\theta)-R_{2}^{-n}w_{k}(\theta)\right] \sin(n\theta) \, \textrm{d}\theta, \\
c_n  =  \dfrac{1}{\pi(R_{1}^{ 2n}-R_{2}^{ 2n})}\int_0^{2\pi} \left[R_{1}^{n}z_{k}(\theta)-R_{2}^{n}w_{k}(\theta)\right] \cos(n\theta) \, \textrm{d}\theta, \\
d_n  =  \dfrac{1}{\pi(R_{1}^{-2n}-R_{2}^{-2n})}\int_0^{2\pi} \left[R_{1}^{-n}z_{k}(\theta)-R_{2}^{-n}w_{k}(\theta)\right] \cos(n\theta) \, \textrm{d}\theta.
\end{gather*}
Since $z_{k}(\theta)$ and $w_{k}(\theta)$ depend on $T_{0}(r,\theta),\hdots,T_{k-1}(r,\theta)$, the calculations are performed sequentially for $k = 1,2,3,\ldots$ where $T_{0}(r)$ is used to calculate $T_{1}(r,\theta)$, $T_{0}(r)$ and $T_{1}(r,\theta)$ are used to calculate $T_{2}(r,\theta)$ and so forth. These calculations are implemented using symbolic computations in MATLAB allowing any number of terms in the perturbation expansion (\ref{eq:ann_T_expansion}) to be evaluated.

\begin{figure}[t]
	\centering
	\fbox{\includegraphics[width=0.8\textwidth]{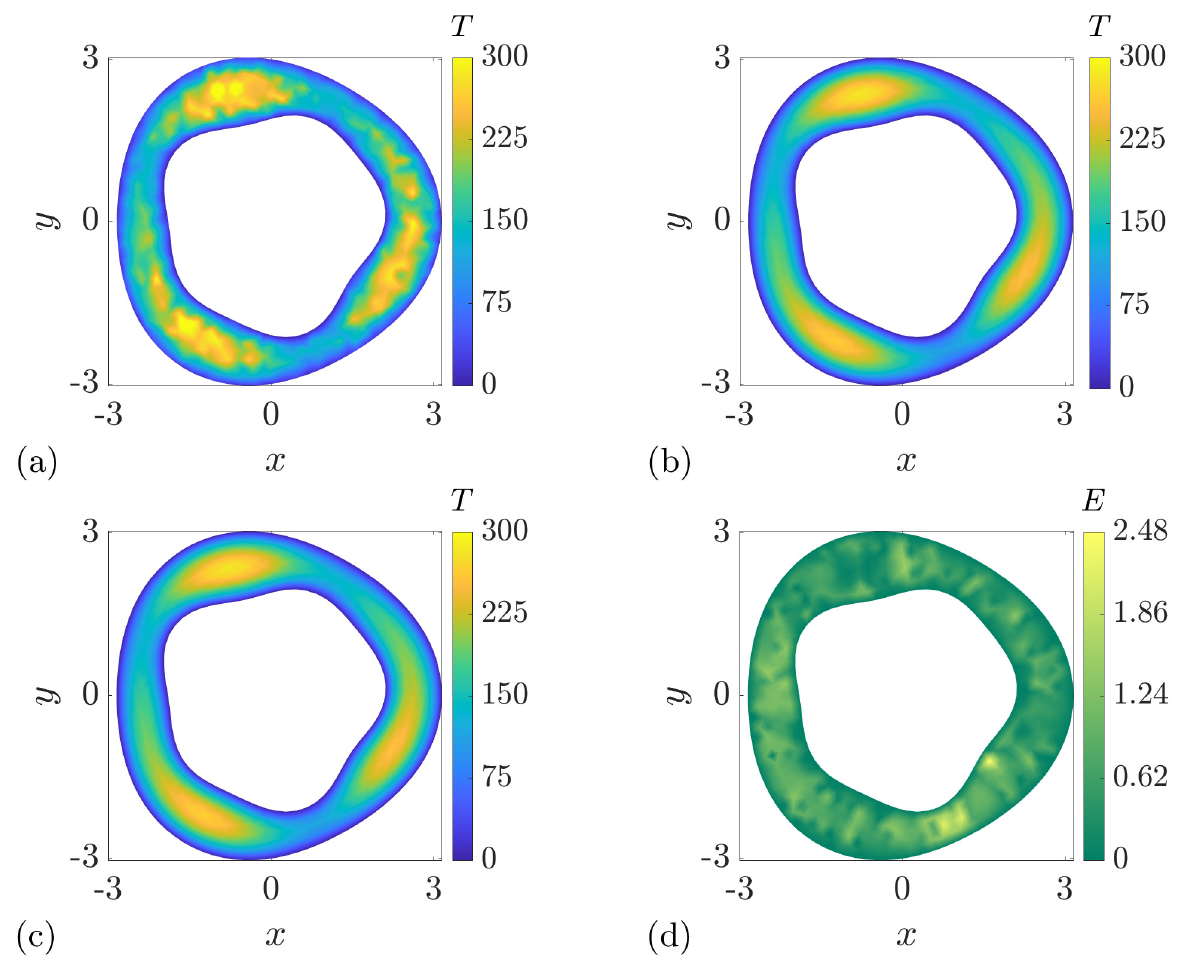}}
		\caption{MET on a perturbed annulus with absorbing inner and outer boundaries. (a) stochastic simulations (b) perturbation solution defined in (\ref{eq:ann_T_expansion}), (\ref{eq:ann1_T0}) and (\ref{eq:ann1_T_fourier}) (c) numerical solution \revision{(d) relative percentage difference between the perturbation and numerical solutions}. All results correspond to the parameter values $R_{1} = 2$, $R_{2} = 3$, $\varepsilon=0.05$, $g_{1}(\theta) = \sin(3\theta) + \cos(5\theta)$, $g_{2}(\theta) = \cos(3\theta)$, $\delta = 0.05$, $\tau = 1$, $P = 1$ and $D = 6.25\times 10^{-4}$. The results in (a) and (c) were produced using a mesh consisting of 573 nodes and 986 triangular elements. The results in (b) were produced using the first three terms in the perturbation expansion (\ref{eq:ann_T_expansion}) and the first 25 terms in the Fourier series (\ref{eq:ann1_T_fourier}).}
	\label{fig:annulus1}
\end{figure}

Results in Figure \ref{fig:annulus1} compare the MET obtained from stochastic simulations to the MET obtained from the perturbation and numerical solutions for a perturbed annulus with absorbing inner and outer boundaries. These results show that the perturbation solution compares very well with the numerical solution and with the stochastic simulations using only the first three terms in the perturbation expansion (\ref{eq:ann_T_expansion}) and the first 25 terms in the Fourier series (\ref{eq:ann1_T_fourier}). We also see that perturbing the inner and outer boundaries of the ellipse produces three local maxima of MET not present in the unperturbed annulus (Figure \ref{fig:annulus1}).  These features are driven by the interplay of the geometry of the inner and outer boundaries, which is interesting since the effect of geometry on exit time is an active area of research~\cite{Wilson2020}.

\subsubsection{Reflecting inner boundary and absorbing outer boundary}
\label{sec:perturbed_annulus2}
We now consider the case of a reflecting inner boundary and absorbing outer boundary where $\nabla T \cdot \mathbf{n}(\theta)=0$ on $r = \mathcal{R}_{1}(\theta)$ and $T = 0$ on $r = \mathcal{R}_{2}(\theta)$:
\begin{gather}
\label{eq:ann2_T_inner}
\nabla T(R_{1} + R_{1} \epsilon g_{1}(\theta), \theta)\cdot\mathbf{n}(\theta) = 0, \\
\label{eq:ann2_T_outer}
T(R_{2} + R_{2} \epsilon g_{2}(\theta), \theta) = 0.
\end{gather}
Here $\mathbf{n}(\theta) = -\mathcal{R}_{1}(\theta)\mathbf{e}_{r} + \mathcal{R}_{1}'(\theta)\mathbf{e}_{\theta} = -R_{1}[1+\varepsilon g_{1}(\theta)]\mathbf{e}_{r} + R_{1}\varepsilon g'_{1}(\theta)\mathbf{e}_{\theta}$ is a vector normal to the curve $r = \mathcal{R}_{1}(\theta)$ with $\mathbf{e}_{r}$ and $\mathbf{e}_{\theta}$ denoting the unit vectors pointing in the positive $r$ and $\theta$ directions, respectively. The absorbing outer boundary condition (\ref{eq:ann2_T_outer}) is treated in the same manner as in Section \ref{sec:perturbed_annulus1} to give the corresponding outer boundary conditions for each term in the perturbation solution (\ref{eq:ann_T_expansion}):
\begin{align}
\mathcal{O}(1):\quad T_{1}(R_{2},\theta) &= 0,\\
\mathcal{O}(\varepsilon^{k}):\quad T_k(R_{2}, \theta) &= w_{k}(\theta),\quad k = 1,2,3,\hdots,
\end{align}
where we define
\begin{gather*}
w_{k}(\theta) = -\sum_{i=1}^{k} \frac{(R_{2}g_{2}(\theta))^i}{i!} \dfrac{\partial^i T_{k-i}}{\partial r^i}(R_{2},\theta).
\end{gather*}
Inserting the definition of $\mathbf{n}(\theta)$ into Equation (\ref{eq:ann2_T_inner}), using the gradient operator in polar coordinates and multiplying both sides by $R_{1}[1+\varepsilon g_{1}(\theta)]$ yields the following equivalent form of the reflecting boundary condition (\ref{eq:ann2_T_inner}):
\begin{gather}
\label{eq:ann2_T_inner2}
-[R_{1}+R_{1}\varepsilon g_{1}(\theta)]^2\dfrac{\partial T}{\partial r}(R_{1}+R_{1}\varepsilon g_{1}(\theta), \theta) + R_{1}\epsilon g_{1}'(\theta)\dfrac{\partial T}{\partial \theta}(R_{1}+R_{1}\varepsilon g_{1}(\theta), \theta) = 0.
\end{gather}
Expanding the derivatives in Equation (\ref{eq:ann2_T_inner2}) in Taylor series centered at $r = R_{1}$, introducing the perturbation expansion (\ref{eq:ann_T_expansion}) and equating powers of $\varepsilon$ yields the appropriate inner boundary conditions for each term in the perturbation solution (\ref{eq:ann_T_expansion}):
\begin{align}
\label{eq:ann2_T_inner_leading}
\mathcal{O}(1):\quad \frac{\partial T_{0}}{\partial r}(R_{1},\theta) &= 0,\\
\label{eq:ann2_T_inner_higher}
\mathcal{O}(\varepsilon^{k}):\quad \frac{\partial T_{k}}{\partial r}(R_{1},\theta) &= z_{k}(\theta),\quad k = 1,2,3,\hdots,
\end{align}
where we define
\begin{gather}
z_{k}(\theta) = \frac{g_1'(\theta)}{R_{1}}\sum_{i=0}^{k-1} \frac{[R_{1}g_1(\theta)]^{i}}{i!}\frac{\partial^{i+1}T_{k-i-1}}{\partial\theta\partial r^{i}}(R_{1},\theta) - \sum_{i=1}^{k} \frac{[R_{1}g_1(\theta)]^{i}}{i!}\frac{\partial^{i+1}T_{k-i}}{\partial r^{i+1}}(R_{1},\theta) \notag \\ - 2g_1(\theta)\sum_{i=0}^{k-1} \frac{[R_{1}g_1(\theta)]^{i}}{i!}\frac{\partial^{i+1}T_{k-i-1}}{\partial r^{i+1}}(R_{1},\theta)
-g_1(\theta)^{2}\sum_{i=0}^{k-2} \frac{[R_{1}g_1(\theta)]^{i}}{i!}\frac{\partial^{i+1}T_{k-i-2}}{\partial r^{i+1}}(R_{1},\theta).
\end{gather}
In summary, we have derived a BVP for each term in the perturbation solution (\ref{eq:ann_T_expansion}). As in the previous section, the first term $T_{0}(r,\theta)\equiv T_{0}(r)$ depends on $r$ only and satisfies the corresponding unperturbed annulus problem discussed in Section \ref{sec:annulus}:
\begin{gather*}
\frac{D}{r}\frac{\text{d}}{\text{d}r}\left(r\frac{\text{d}T_0}{\text{d}r}\right) = -1,  \quad R_{1}<r<R_{2},\\
\dfrac{\textrm{d}T_0}{\textrm{d}r}(R_{1}) = 0, \quad
T_0(R_{2}) = 0,
\end{gather*}
which has solution (\ref{eq:a_sol2}):
\begin{gather}
\label{eq:ann2_T0}
T_{0}(r) = \frac{R_{2}^{2}-r^{2}}{4D} + \frac{R_{1}^{2}}{2D}\log(r/R_{2}).
\end{gather}
In contrast, the higher-order terms $T_{k}(r,\theta)$ ($k = 1,2,3,\hdots)$ in the perturbation solution (\ref{eq:ann_T_expansion}) depend on both $r$ and $\theta$ and satisfy
\begin{gather}
\label{eq:ann2_Tk_pde}
\nabla^2 T_{k}= 0, \quad R_{1}<r<R_{2}, \\
\label{eq:ann2_Tk_bcs}
\dfrac{\partial T_{k}}{\partial r}(R_{1}, \theta) = z_{k}(\theta), \quad
T_{k}(R_{2}, \theta) = w_{k}(\theta).
\end{gather}
Again, each higher-order correction term is given by the solution of Laplace's equation on the unperturbed annulus with the boundary functions $w_{k}(\theta)$ and $z_{k}(\theta)$ accounting for the projection of the boundary conditions onto the unperturbed annulus domain. Solving the BVP (\ref{eq:ann2_Tk_pde})--(\ref{eq:ann2_Tk_bcs}) using separation of variables gives:
\begin{equation}
\label{eq:ann2_T_fourier}
T_k(r, \theta) = c_0 + d_0 \log r + \sum_{n=1}^\infty \left[\left(c_n r^n + d_n r^{-n}\right) \cos(n\theta) + \left(a_n r^n + b_n r^{-n}\right) \sin(n\theta)\right],
\end{equation}
for $k = 1,2,3,\hdots$, where
\begin{gather}
c_0  =  \dfrac{1}{2\pi}\int_0^{2\pi} w_{k}(\theta) -R_{1}\log(R_{2})z_{k}(\theta) \, \textrm{d}\theta, \\
d_0  =  \dfrac{1}{2\pi}\int_0^{2\pi} R_{1} z_{k}(\theta) \, \textrm{d}\theta, \\
a_n  = \dfrac{1}{n \pi\mathcal{V}_n}\int_0^{2\pi} \left[nR_{1}^{-n-1} w_{k}(\theta) + R_{2}^{-n} z_{k}(\theta)\right] \sin(n\theta) \, \textrm{d}\theta, \\
b_n  =  \dfrac{1}{n \pi\mathcal{V}_n}\int_0^{2\pi} \left[nR_{1}^{n-1} w_{k}(\theta) - R_{2}^n z_{k}(\theta)\right] \sin(n\theta)  \, \textrm{d}\theta, \\
c_n  =  \dfrac{1}{n\pi\mathcal{V}_n}\int_0^{2\pi} \left[nR_{1}^{-n-1} w_{k}(\theta) + R_{2}^{-n}z_{k}(\theta)\right] \cos(n\theta)  \, \textrm{d}\theta,\\
d_n  =  \dfrac{1}{n\pi\mathcal{V}_n}\int_0^{2\pi} \left[nR_{1}^{n-1}  w_{k}(\theta) - R_{2}^n z_{k}(\theta)\right] \cos(n\theta)  \, \textrm{d}\theta, \\
\label{eq:ann2_T_fourier_Vn}
\mathcal{V}_n =  R_{1}^{-n-1}R_{2}^n + R_{1}^{n-1}R_{2}^{-n}.
\end{gather}

\begin{figure}[t]
	\centering
	\fbox{\includegraphics[width=0.8\textwidth]{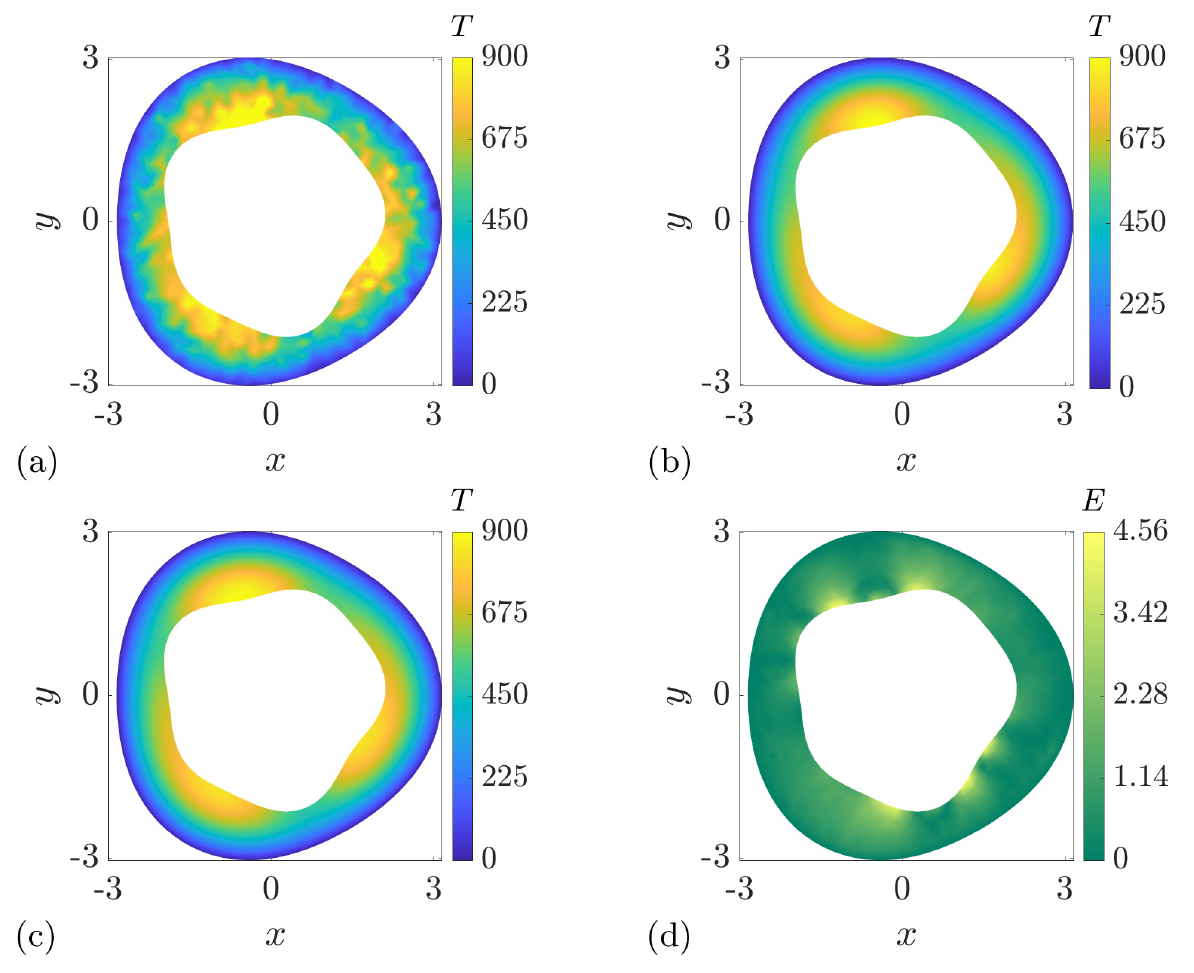}}
	\caption{MET on a perturbed annulus with a reflecting inner boundary and absorbing outer boundary. (a) stochastic simulations (b) perturbation solution defined in equations (\ref{eq:ann_T_expansion}), (\ref{eq:ann2_T0}) and (\ref{eq:ann2_T_fourier})  (c) numerical solution \revision{(d) relative percentage difference between the perturbation and numerical solutions}. All results correspond to the parameter values $R_{1} = 2$, $R_{2} = 3$, $\varepsilon=0.05$, $g_{1}(\theta) = \sin(3\theta) + \cos(5\theta)$, $g_{2}(\theta) = \cos(3\theta)$, $\delta = 0.05$, $\tau = 1$, $P = 1$ and $D = 6.25\times 10^{-4}$. The results in (a) and (c) were produced using a mesh consisting of 573 nodes and 986 triangular elements. The results in (b) were produced using the three terms in the perturbation expansion (\ref{eq:ann_T_expansion}) and the first 25 terms in the Fourier series (\ref{eq:ann2_T_fourier}).}
	\label{fig:annulus2}
\end{figure}

Results in Figure \ref{fig:annulus2} compare the MET obtained from stochastic simulations to the MET obtained from the perturbation and numerical solutions for a perturbed annulus with reflecting inner boundary and absorbing outer boundary. As in Figure \ref{fig:annulus1} these results show that an accurate approximation can be obtained using only the first three terms in the perturbation expansion (\ref{eq:ann_T_expansion}) and the first 25 terms in the Fourier series (\ref{eq:ann2_T_fourier}). The three local maxima of MET evident in Figure \ref{fig:annulus3} are now located on the inner boundary.

\revision{Note that the perturbed annulus with a reflecting inner boundary and an absorbing outer boundary collapses to a perturbed disc with an absorbing outer boundary when $R_{1} = 0$. Making this substitution in (\ref{eq:ann2_T0}) yields $T_{0}(r) = (R_{2}^{2}-r^{2})/(4D)$ while taking the limit of (\ref{eq:ann2_T_fourier})--(\ref{eq:ann2_T_fourier_Vn}) as $R_{1}$ tends to zero yields $T_{k}(r,\theta) = c_{0} + \sum_{n=1}^{\infty}\left[c_{n}r^{n}\cos(n\theta) + a_{n}r^{n}\sin(n\theta)\right]$, where
\begin{gather*}
c_{0} = \dfrac{1}{2\pi}\int_0^{2\pi} w_{k}(\theta)\, \textrm{d}\theta,\quad c_n  =\dfrac{1}{\pi R_{2}^n}\int_0^{2\pi}  w_{k}(\theta) \cos(n\theta)  \, \textrm{d}\theta,\quad a_n  = \dfrac{1}{\pi R_{2}^{n}}\int_0^{2\pi} w_{k}(\theta)\sin(n\theta) \, \textrm{d}\theta.
\end{gather*}
As expected, this simplification recovers the perturbation solution developed in our previous work \cite{Simpson2021} for a perturbed disc with absorbing outer boundary.}

\revision{\subsubsection{Absorbing inner boundary and reflecting outer boundary}
\label{sec:perturbed_annulus3}
We now consider the third and final case for the annulus where the inner boundary is absorbing and the outer boundary boundary reflecting. Here $T = 0$ on $r = \mathcal{R}_{1}(\theta)$ and $\nabla T \cdot \mathbf{n}(\theta)=0$ on $r = \mathcal{R}_{2}(\theta)$:
\begin{gather*}
\label{eq:ann3_T_inner}
T(R_{1} + R_{1} \epsilon g_{1}(\theta), \theta) = 0, \\
\label{eq:ann3_T_outer}
\nabla T(R_{2} + R_{2} \epsilon g_{2}(\theta), \theta)\cdot\mathbf{n}(\theta) = 0,
\end{gather*}
where $\mathbf{n}(\theta) = -\mathcal{R}_{2}(\theta)\mathbf{e}_{r} + \mathcal{R}_{2}'(\theta)\mathbf{e}_{\theta} = -R_{2}[1+\varepsilon g_{2}(\theta)]\mathbf{e}_{r} + R_{2}\varepsilon g'_{2}(\theta)\mathbf{e}_{\theta}$ is a vector normal to the curve $r = \mathcal{R}_{2}(\theta)$. Developing the perturbation solution for this case proceeds similarly to the working carried out in Sections \ref{sec:perturbed_annulus1} and \ref{sec:perturbed_annulus2} with each term in the perturbation solution (\ref{eq:ann_T_expansion}) satisfying a BVP on the unperturbed annulus domain. The first term $T_{0}(r,\theta)\equiv T_{0}(r)$ satisfies:
\begin{gather*}
\frac{D}{r}\frac{\text{d}}{\text{d}r}\left(r\frac{\text{d}T_0}{\text{d}r}\right) = -1,  \quad R_{1}<r<R_{2},\\
T_0(R_{1}) = 0, \quad
\dfrac{\textrm{d}T_0}{\textrm{d}r}(R_{2}) = 0,
\end{gather*}
while the higher-order terms $T_{k}(r,\theta)$ ($k = 1,2,3,\hdots)$ in the perturbation solution (\ref{eq:ann_T_expansion}) satisfy
\begin{gather*}
\nabla^2 T_{k}= 0, \quad R_{1}<r<R_{2}, \\
T_{k}(R_{1}, \theta) = z_{k}(\theta), \quad
\dfrac{\partial T_{k}}{\partial r}(R_{2}, \theta) = w_{k}(\theta),
\end{gather*}
where $z_{k}(\theta)$ and $w_{k}(\theta)$ are defined as:
\begin{gather*}
z_{k}(\theta) = -\sum_{i=1}^{k} \frac{(R_{1}g_{1}(\theta))^i}{i!} \dfrac{\partial^i T_{k-i}}{\partial r^i}(R_{1},\theta),\\
w_{k}(\theta) = \frac{g_2'(\theta)}{R_{2}}\sum_{i=0}^{k-1} \frac{[R_{2}g_2(\theta)]^{i}}{i!}\frac{\partial^{i+1}T_{k-i-1}}{\partial\theta\partial r^{i}}(R_{2},\theta) - \sum_{i=1}^{k} \frac{[R_{2}g_2(\theta)]^{i}}{i!}\frac{\partial^{i+1}T_{k-i}}{\partial r^{i+1}}(R_{2},\theta) \notag \\ - 2g_2(\theta)\sum_{i=0}^{k-1} \frac{[R_{2}g_2(\theta)]^{i}}{i!}\frac{\partial^{i+1}T_{k-i-1}}{\partial r^{i+1}}(R_{2},\theta)
-g_2(\theta)^{2}\sum_{i=0}^{k-2} \frac{[R_{2}g_2(\theta)]^{i}}{i!}\frac{\partial^{i+1}T_{k-i-2}}{\partial r^{i+1}}(R_{2},\theta).
\end{gather*}

\begin{figure}[t]
	\centering
	\fbox{\includegraphics[width=0.8\textwidth]{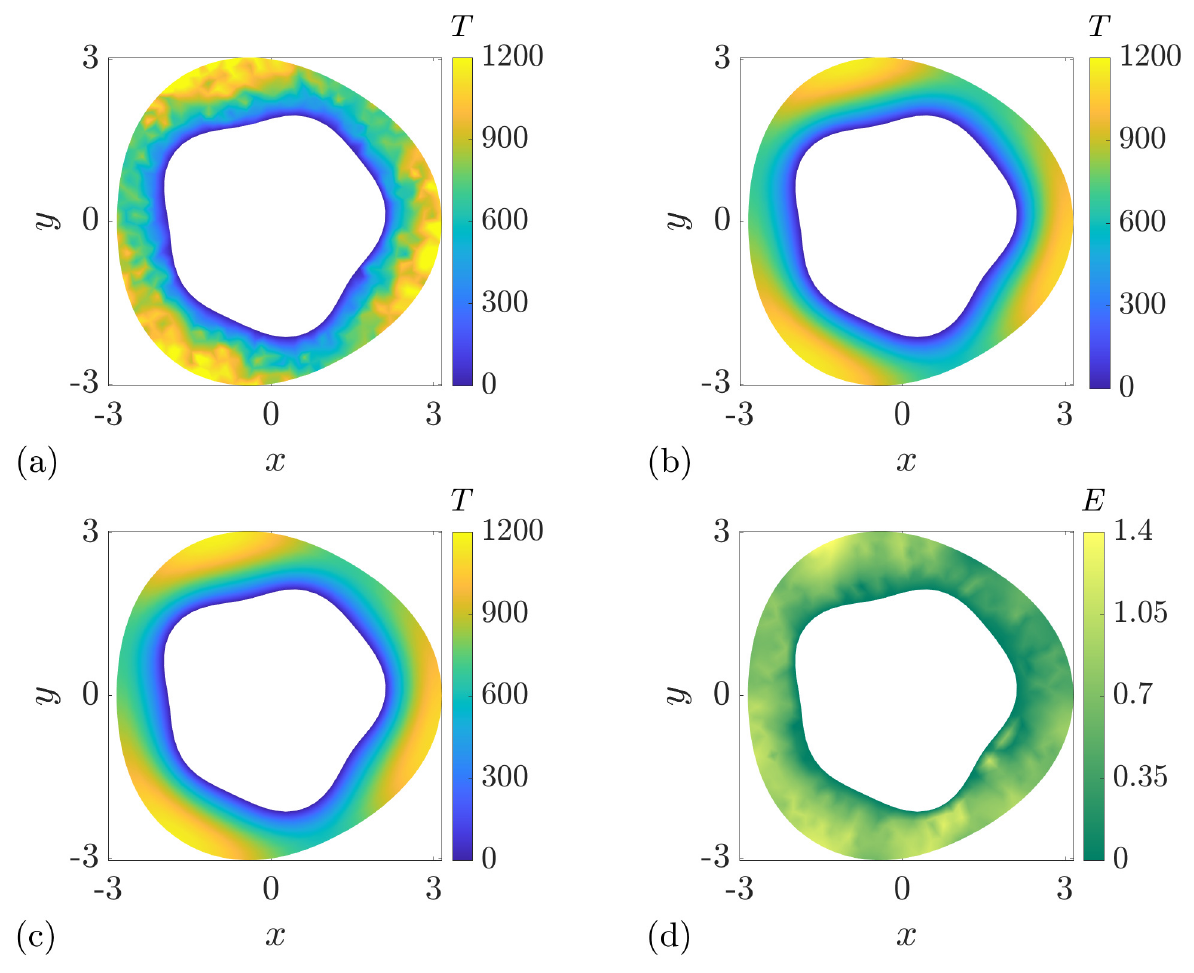}}
	\caption{MET on a perturbed annulus with an absorbing inner boundary and reflecting outer boundary. (a) stochastic simulations (b) perturbation solution defined in equations (\ref{eq:ann_T_expansion}), (\ref{eq:ann3_T0}) and (\ref{eq:ann3_T_fourier}) (c) numerical solution \revision{(d) relative percentage difference between the perturbation and numerical solutions}. All results correspond to the parameter values $R_{1} = 2$, $R_{2} = 3$, $\varepsilon=0.05$, $g_{1}(\theta) = \sin(3\theta) + \cos(5\theta)$, $g_{2}(\theta) = \cos(3\theta)$, $\delta = 0.05$, $\tau = 1$, $P = 1$ and $D = 6.25\times 10^{-4}$. The results in (a) and (c) were produced using a mesh consisting of 573 nodes and 986 triangular elements. The results in (b) were produced using the three terms in the perturbation expansion (\ref{eq:ann_T_expansion}) and the first 25 terms in the Fourier series (\ref{eq:ann3_T_fourier}).}
	\label{fig:annulus3}
\end{figure}

\noindent Solving these BVPs yields the leading-order term:
\begin{gather}
\label{eq:ann3_T0}
T_{0}(r) = \frac{R_{1}^{2}-r^{2}}{4D} + \frac{R_{2}^{2}}{2D}\log(r/R_{1}),
\end{gather}
and the higher-order terms:
\begin{gather}
\label{eq:ann3_T_fourier}
T_k(r, \theta) = c_0 + d_0 \log r + \sum_{n=1}^\infty \left[\left(c_n r^n + d_n r^{-n}\right) \cos(n\theta) + \left(a_n r^n + b_n r^{-n}\right) \sin(n\theta)\right],
\end{gather}
for $k = 1,2,3,\hdots$, where
\begin{gather*}
c_0  =  \dfrac{1}{2\pi}\int_0^{2\pi} z_{k}(\theta) -R_{2}\log(R_{1})w_{k}(\theta) \, \textrm{d}\theta, \\
d_0  =  \dfrac{1}{2\pi}\int_0^{2\pi} R_{2} w_{k}(\theta) \, \textrm{d}\theta, \\
a_n  = \dfrac{1}{n \pi\mathcal{V}_n}\int_0^{2\pi} \left[nR_{2}^{-n-1} z_{k}(\theta) + R_{1}^{-n} w_{k}(\theta)\right] \sin(n\theta) \, \textrm{d}\theta, \\
b_n  =  \dfrac{1}{n \pi\mathcal{V}_n}\int_0^{2\pi} \left[nR_{2}^{n-1} z_{k}(\theta) - R_{1}^n w_{k}(\theta)\right] \sin(n\theta)  \, \textrm{d}\theta, \\
c_n  =  \dfrac{1}{n\pi\mathcal{V}_n}\int_0^{2\pi} \left[nR_{2}^{-n-1} z_{k}(\theta) + R_{1}^{-n}w_{k}(\theta)\right] \cos(n\theta)  \, \textrm{d}\theta,\\
d_n  =  \dfrac{1}{n\pi\mathcal{V}_n}\int_0^{2\pi} \left[nR_{2}^{n-1}  z_{k}(\theta) - R_{1}^n w_{k}(\theta)\right] \cos(n\theta)  \, \textrm{d}\theta, \\
\mathcal{V}_n =  R_{2}^{-n-1}R_{1}^n + R_{2}^{n-1}R_{1}^{-n}.
\end{gather*}
Results in Figure \ref{fig:annulus3} show that the perturbation solution for this case remains accurate and that the three local maxima of MET evident in Figures \ref{fig:annulus1} and \ref{fig:annulus2} are now located on the outer boundary.
}

\subsection{Compound disc}
\label{sec:compound_disc}
We now review the case of a compound disc where $\Omega = \{0 < r < R_{1}\}\cup \{R_{1} < r < R_{2}\}$ and $P$ (and hence $D$) are piecewise constant across $\Omega$:
\begin{gather*}
P = \begin{cases} P_{1}, & 0 < r < R_{1},\\ P_{2}, & R_{1} < r < R_{2}, \end{cases}\quad D = \begin{cases} D_{1}, & 0 < r < R_{1},\\ D_{2}, & R_{1} < r < R_{2}, \end{cases}
\end{gather*}
where $D_{1} = P_{1}\delta^2 / (4 \tau)$ and $D_{2} = P_{2}\delta^2 / (4 \tau)$. Here the inner region ($0 < r < R_{1}$) and outer region ($R_{1} < r < R_{2}$) of the disc are separated by an interface at $r = R_{1}$ and an absorbing boundary condition is applied at the outer boundary at $r = R_{2}$. \revision{To be consistent with our treatment of the interface in the stochastic simulations (Appendix \ref{app:stochastic_model}), both the MET and diffusive flux are taken to continuous at the interface \cite{Carr2020}}. For this unperturbed problem the MET can be expressed in terms of the radial coordinate, $r$, only:
\begin{gather}
\label{eq:cd_ode1}
\frac{D_{1}}{r}\frac{\text{d}}{\text{d}r}\left(r\frac{\text{d}T^{(1)}}{\text{d}r}\right) = -1,\quad 0 < r < R_{1},\\
\label{eq:cd_ode2}
\frac{D_{2}}{r}\frac{\text{d}}{\text{d}r}\left(r\frac{\text{d}T^{(2)}}{\text{d}r}\right) = -1,\quad R_{1} < r < R_{2},\\
\label{eq:cd_ics}
T^{(1)}(R_{1}) = T^{(2)}(R_{1}),\quad D_{1}\frac{\text{d}T^{(1)}}{\text{d}r}(R_{1}) = D_{2}\frac{\text{d}T^{(2)}}{\text{d}r}(R_{1}),\\
\label{eq:cd_bcs}
\frac{\text{d}T^{(1)}}{\text{d}r}(0) = 0,\quad T^{(2)}(R_{2}) = 0,
\end{gather}
where $T^{(1)}(r)$ and $T^{(2)}(r)$ denote the MET for $0 < r < R_{1}$ and $R_{1}<r<R_{2}$, respectively. The solution of the BVP (\ref{eq:cd_ode1})--(\ref{eq:cd_bcs}) is given by \cite{Carr2020}:
\begin{gather}
\label{eq:cd_exact_S}
T^{(1)}(r) = \frac{R_{1}^{2}-r^{2}}{4D_{1}} + \frac{R_{2}^{2}-R_{1}^{2}}{4D_{2}},\\
\label{eq:cd_exact_T}
T^{(2)}(r) = \frac{R_{2}^{2}-r^{2}}{4D_{2}}.
\end{gather}

\subsection{Compound disc with a perturbed interface} \label{sec:perturbed_compound_disc}
We now consider the case of a heterogeneous compound disc with a perturbed interface where $\Omega = \{0 < r < \mathcal{R}_{1}(\theta)\}\cup \{\mathcal{R}_{1}(\theta) < r < R_{2}\}$ and
\begin{gather*}
P = \begin{cases} P_{1}, & 0 < r < \mathcal{R}_{1}(\theta),\\ P_{2}, & \mathcal{R}_{1}(\theta) < r < R_{2}, \end{cases}\quad D = \begin{cases} D_{1}, & 0 < r < \mathcal{R}_{1}(\theta),\\ D_{2}, & \mathcal{R}_{1}(\theta) < r < R_{2}. \end{cases}
\end{gather*}
Here the inner region ($0 < r < \mathcal{R}_{1}(\theta)$) and outer region ($\mathcal{R}_{1}(\theta) < r < R_{2}$) of the compound disc are separated by a perturbed interface described by the polar curve $r = \mathcal{R}_{1}(\theta)$, where $\mathcal{R}_{1}(\theta) = R_{1}(1+\varepsilon g(\theta))$. In a similar manner to Section \ref{sec:perturbed_annulus}, $\varepsilon\ll 1$ is the perturbation parameter, $R_{1}$ is the radius of the unperturbed interface and $g(\theta)$ is a smooth $\mathcal{O}(1)$ periodic function with period $2\pi$. In this case, the MET satisfies the BVP:
\begin{gather}
\label{eq:pcd_pde1}
D_{1}\nabla^{2}T^{(1)} = -1,\quad 0 < r < \mathcal{R}_{1}(\theta),\\
\label{eq:pcd_pde2}
D_{2}\nabla^{2}T^{(2)} = -1,\quad \mathcal{R}_{1}(\theta) < r < R_{2},\\
\label{eq:pcd_ic1}
T^{(1)}(\mathcal{R}_{1}(\theta),\theta) = T^{(2)}(\mathcal{R}_{1}(\theta),\theta),\\
\label{eq:pcd_ic2}
D_{1}\nabla T^{(1)}(\mathcal{R}_{1}(\theta),\theta)\cdot\mathbf{n}(\theta) = D_{2}\nabla T^{(2)}(\mathcal{R}_{1}(\theta),\theta)\cdot\mathbf{n}(\theta),\\
\label{eq:pcd_bc}
T^{(2)}(R_{2},\theta) = 0,
\end{gather}
where $T^{(1)}(r,\theta)$ and $T^{(2)}(r,\theta)$ denote the MET for $0 < r < \mathcal{R}_{1}(\theta)$ and $\mathcal{R}_{1}(\theta)<r<R_{2}$, respectively, and
$\mathbf{n}(\theta) = -R_{1}[1+\varepsilon g(\theta)]\mathbf{e}_{r} + R_{1}\varepsilon g'(\theta)\mathbf{e}_{\theta}$ is a vector normal to the curve $r = \mathcal{R}_{1}(\theta)$. We assume the solutions $T^{(1)}(r,\theta)$ and $T^{(2)}(r,\theta)$ of equations (\ref{eq:pcd_pde1})--(\ref{eq:pcd_bc}) can be expanded in powers of $\varepsilon$:
\begin{align}
\label{eq:S_expansion}
T^{(1)}(r,\theta) = \sum_{k=0}^{\infty}\varepsilon^{k}T^{(1)}_{k}(r,\theta),\\
\label{eq:T_expansion}
T^{(2)}(r,\theta) = \sum_{k=0}^{\infty}\varepsilon^{k}T^{(2)}_{k}(r,\theta),
\end{align}
where $T_0^{(1)}(r, \theta)$ and $T_{0}^{(2)}(r,\theta)$ satisfy the BVP (\ref{eq:pcd_pde1})--(\ref{eq:pcd_bc}) on the unperturbed compound disc domain and $T_{k}^{(1)}(r,\theta)$ and $T_{k}^{(2)}(r,\theta)$ ($k = 1,2,3,\hdots$) satisfy Laplace's equation subject to boundary conditions (\ref{eq:pcd_ic1})--(\ref{eq:pcd_bc}) on the unperturbed compound disc domain.

Substituting equations (\ref{eq:S_expansion}) and (\ref{eq:T_expansion}) into equation (\ref{eq:pcd_ic1}), expanding $T^{(1)}_{k}(r,\theta)$ and $T^{(2)}_{k}(r,\theta)$ at $r = R_{1}(1+\varepsilon g(\theta))$ in Taylor series centered at $x = R_{1}$ and equating powers of $\varepsilon$ yields the first set of interface conditions for the terms in the perturbation expansions (\ref{eq:S_expansion}) and (\ref{eq:T_expansion}):
\begin{align*}
\mathcal{O}(1):\quad T^{(1)}_{0}(R_{1},\theta) &= T^{(2)}_{0}(R_{1},\theta),\\
\mathcal{O}(\varepsilon^{k}):\quad T^{(1)}_{k}(R_{1},\theta) &= T^{(2)}_{k}(R_{1},\theta) + w_{k}(\theta),\quad k = 1,2,3,\hdots,
\end{align*}
where we have defined
\begin{gather}
\label{eq:wk}
w_{k}(\theta) = \sum_{i=1}^{k} \frac{[R_{1}g(\theta)]^{i}}{i!}\left[\frac{\partial^{i}T^{(2)}_{k-i}}{\partial r^{i}}(R_{1},\theta) - \frac{\partial^{i}T^{(1)}_{k-i}}{\partial r^{i}}(R_{1},\theta)\right].
\end{gather}
Evaluating the dot product in equation (\ref{eq:pcd_ic2}) using both the definition of $\mathbf{n}(\theta)$ and the form of the gradient operator in polar coordinates gives
\begin{multline}
\label{eq:pcd_ic2_2}
-D_{1}[R_{1}+R_{1}\varepsilon g(\theta)]^{2}\frac{\partial T^{(1)}}{\partial r}(R_{1}+R_{1}\varepsilon g(\theta),\theta) + D_{1}R_{1}\varepsilon g'(\theta)\frac{\partial T^{(1)}}{\partial \theta}(R_{1}+R_{1}\varepsilon g(\theta),\theta)\\
= -D_{2}[R_{1}+R_{1}\varepsilon g(\theta)]^{2}\frac{\partial T^{(2)}}{\partial r}(R_{1}+R_{1}\varepsilon g(\theta),\theta) + D_{2}R_{1}\varepsilon g'(\theta)\frac{\partial T^{(2)}}{\partial \theta}(R_{1}+R_{1}\varepsilon g(\theta),\theta).
\end{multline}
Substituting equations (\ref{eq:S_expansion}) and (\ref{eq:T_expansion}) into equation (\ref{eq:pcd_ic2_2}), expanding the derivatives of $T^{(1)}_{k}(r,\theta)$ and $T^{(2)}_{k}(r,\theta)$ at $r = R_{1}(1+\varepsilon g(\theta))$ in Taylor series centered at $x = R_{1}$ and equating powers of $\varepsilon$ yields the second set of interface conditions for the terms in the perturbation expansions (\ref{eq:S_expansion}) and (\ref{eq:T_expansion}):
\begin{align*}
\mathcal{O}(1):\quad D_{1}\frac{\partial T^{(1)}_{0}}{\partial r}(R_{1},\theta) &= D_{2}\frac{\partial T^{(2)}_{0}}{\partial r}(R_{1},\theta),\\
\mathcal{O}(\varepsilon^{k}):\quad D_{1}\frac{\partial T^{(1)}_{k}}{\partial r}(R_{1},\theta) &= D_{2}\frac{\partial T^{(2)}_{k}}{\partial r}(R_{1},\theta) + z_{k}(\theta),\quad k = 1,2,3,\hdots,
\end{align*}
where
\begin{align}
\nonumber
z_{k}(\theta) &= \sum_{i=1}^{k} \frac{[R_{1}g(\theta)]^{i}}{i!}\left[D_{2}\frac{\partial^{i+1}T^{(2)}_{k-i}}{\partial r^{i+1}}(R_{1},\theta) - D_{1}\frac{\partial^{i+1}T^{(1)}_{k-i}}{\partial r^{i+1}}(R_{1},\theta)\right]\\
\nonumber
&\qquad + 2g(\theta)\sum_{i=0}^{k-1} \frac{[R_{1}g(\theta)]^{i}}{i!}\left[D_{2}\frac{\partial^{i+1}T^{(2)}_{k-i-1}}{\partial r^{i+1}}(R_{1},\theta) - D_{1}\frac{\partial^{i+1}T^{(1)}_{k-i-1}}{\partial r^{i+1}}(R_{1},\theta)\right]\\
\nonumber
&\qquad + g(\theta)^{2}\sum_{i=0}^{k-2} \frac{[R_{1}g(\theta)]^{i}}{i!}\left[D_{2}\frac{\partial^{i+1}T^{(2)}_{k-i-2}}{\partial r^{i+1}}(R_{1},\theta) - D_{1}\frac{\partial^{i+1}T^{(1)}_{k-i-2}}{\partial r^{i+1}}(R_{1},\theta)\right]\\
\label{eq:zk}
&\qquad - \frac{g'(\theta)}{R_{1}}\sum_{i=0}^{k-1} \frac{[R_{1}g(\theta)]^{i}}{i!}\left[D_{2}\frac{\partial^{i+1}T^{(2)}_{k-i-1}}{\partial\theta\partial r^{i}}(R_{1},\theta) - D_{1}\frac{\partial^{i+1}T^{(1)}_{k-i-1}}{\partial\theta\partial r^{i}}(R_{1},\theta)\right]\!.
\end{align}
In summary, $T^{(1)}_{0}(r,\theta)$ and $T_{0}(r,\theta)$ satisfy the following BVP:
\begin{gather}
\label{eq:pcd_pde1_order0}
D_{1}\nabla^{2}T^{(1)}_{0} = -1,\quad 0 < r < R_{1},\\
D_{2}\nabla^{2}T^{(2)}_{0} = -1,\quad R_{1} < r < R_{2},\\
T^{(1)}_{0}(R_{1},\theta) = T^{(2)}_{0}(R_{1},\theta),\\
D_{1}\frac{\partial T^{(1)}_{0}}{\partial r}(R_{1},\theta) = D_{2}\frac{\partial T^{(2)}_{0}}{\partial r}(R_{1},\theta),\\
\label{eq:pcd_bc_order0}
T^{(2)}_{0}(R_{2},\theta) = 0,
\end{gather}
while $T^{(1)}_{k}(r,\theta)$ and $T^{(2)}_{k}(r,\theta)$ for $k = 1,2,\hdots$, satisfy the following BVP:
\begin{gather}
\label{eq:pcd_pde1_orderk}
D_{1}\nabla^{2}T^{(1)}_{k} = 0,\quad 0 < r < R_{1},\\
D_{2}\nabla^{2}T^{(2)}_{k} = 0,\quad R_{1} < r < R_{2},\\
T^{(1)}_{k}(R_{1},\theta) = T^{(2)}_{k}(R_{1},\theta) + w_{k}(\theta),\\
D_{1}\frac{\partial T^{(1)}_{k}}{\partial r}(R_{1},\theta) = D_{2}\frac{\partial T^{(2)}_{k}}{\partial r}(R_{1},\theta) + z_{k}(\theta),\\
\label{eq:pcd_bc_orderk}
T^{(2)}_{k}(R_{2},\theta) = 0,
\end{gather}
where $w_{k}(\theta)$ and $z_{k}(\theta)$ are given in equations (\ref{eq:wk}) and (\ref{eq:zk}). The BVP (\ref{eq:pcd_pde1_order0})--(\ref{eq:pcd_bc_order0}) is the same as the BVP for the unperturbed compound disc (\ref{eq:cd_ode1})--(\ref{eq:cd_bcs}) and hence has solution:
\begin{gather}
\label{eq:T10}
T^{(1)}_{0}(r) = \frac{R_{1}^{2}-r^{2}}{4D_{1}} + \frac{R_{2}^{2}-R_{1}^{2}}{4D_{2}},\\
\label{eq:T20}
T^{(2)}_{0}(r) = \frac{R_{2}^{2}-r^{2}}{4D_{2}}.
\end{gather}
The BVP (\ref{eq:pcd_pde1_orderk})--(\ref{eq:pcd_bc_orderk}) can be solved using separation of variables and eigenfunction expansion yielding
\begin{gather}
\label{eq:T1_fourier}
T^{(1)}_{k}(r,\theta) = \sum_{n=1}^{\infty} \left[a_{n}r^{n}\cos(n\theta) + b_{n}r^{n}\sin(n\theta)\right],\\
\label{eq:T2_fourier}
T^{(2)}_{k}(r,\theta) = c_{0}+d_{0}\log(r) + \sum_{n=1}^{\infty}\left[(c_{n}r^{n}+d_{n}r^{-n})\cos(n\theta) + (e_{n}r^{n}+f_{n}r^{-n})\sin(n\theta)\right],
\end{gather}
for $k = 1,2,3,\hdots$, where
\begin{gather*}
a_{n} = \frac{1}{n\pi D_{1}R_{1}^{n-1}}\left[\alpha_{n} + \int_{-\pi}^{\pi}z_{k}(\theta)\cos(n\theta)\,\mathrm{d}\theta\right]\!,\\
b_{n} = \frac{1}{n\pi D_{1}R_{1}^{n-1}}\left[\beta_{n} + \int_{-\pi}^{\pi}z_{k}(\theta)\sin(n\theta)\,\mathrm{d}\theta\right]\!,\\
c_{0} = \frac{-R_{1}\log(R_{2})}{2\pi D_{2}}\alpha_{0},\quad
d_{0} = \frac{R_{1}}{2\pi D_{2}}\alpha_{0},\\
c_{n} = \frac{R_{1}^{n+1}}{n\pi D_{2}(R_{1}^{2n}+R_{2}^{2n})}\alpha_{n},\quad
d_{n} = \frac{-R_{2}^{2n}R_{1}^{n+1}}{n\pi D_{2}(R_{1}^{2n}+R_{2}^{2n})}\alpha_{n},\\
e_{n} = \frac{R_{1}^{n+1}}{n\pi D_{2}(R_{1}^{2n}+R_{2}^{2n})}\beta_{n},\quad
f_{n} = \frac{-R_{2}^{2n}R_{1}^{n+1}}{n\pi D_{2}(R_{1}^{2n}+R_{2}^{2n})}\beta_{n},\\
\alpha_{0} =  \frac{D_{2}}{R_{1}\log(\frac{R_{1}}{R_{2}})}\int_{-\pi}^{\pi}w_{k}(\theta)\,\mathrm{d}\theta,\quad \gamma_{n} = \frac{nD_{1}D_{2}(R_{1}^{2n}+R_{2}^{2n})}{R_{1}\left[(D_{2}-D_{1})R_{1}^{2n} + (D_{2}+D_{1})R_{2}^{2n}\right]},\\
\alpha_{n} = \gamma_{n}\left[\int_{-\pi}^{\pi}w_{k}(\theta)\cos(n\theta)\,\mathrm{d}\theta - \frac{R_{1}}{nD_{1}}\int_{-\pi}^{\pi}z_{k}(\theta)\cos(n\theta)\,\mathrm{d}\theta\right]\!,\\
\beta_{n} = \gamma_{n}\left[\int_{-\pi}^{\pi}w_{k}(\theta)\sin(n\theta)\,\mathrm{d}\theta - \frac{R_{1}}{nD_{1}}\int_{-\pi}^{\pi}z_{k}(\theta)\sin(n\theta)\,\mathrm{d}\theta\right]\!.
\end{gather*}

\begin{figure}[t]
	\centering
	\fbox{\includegraphics[width=0.8\textwidth]{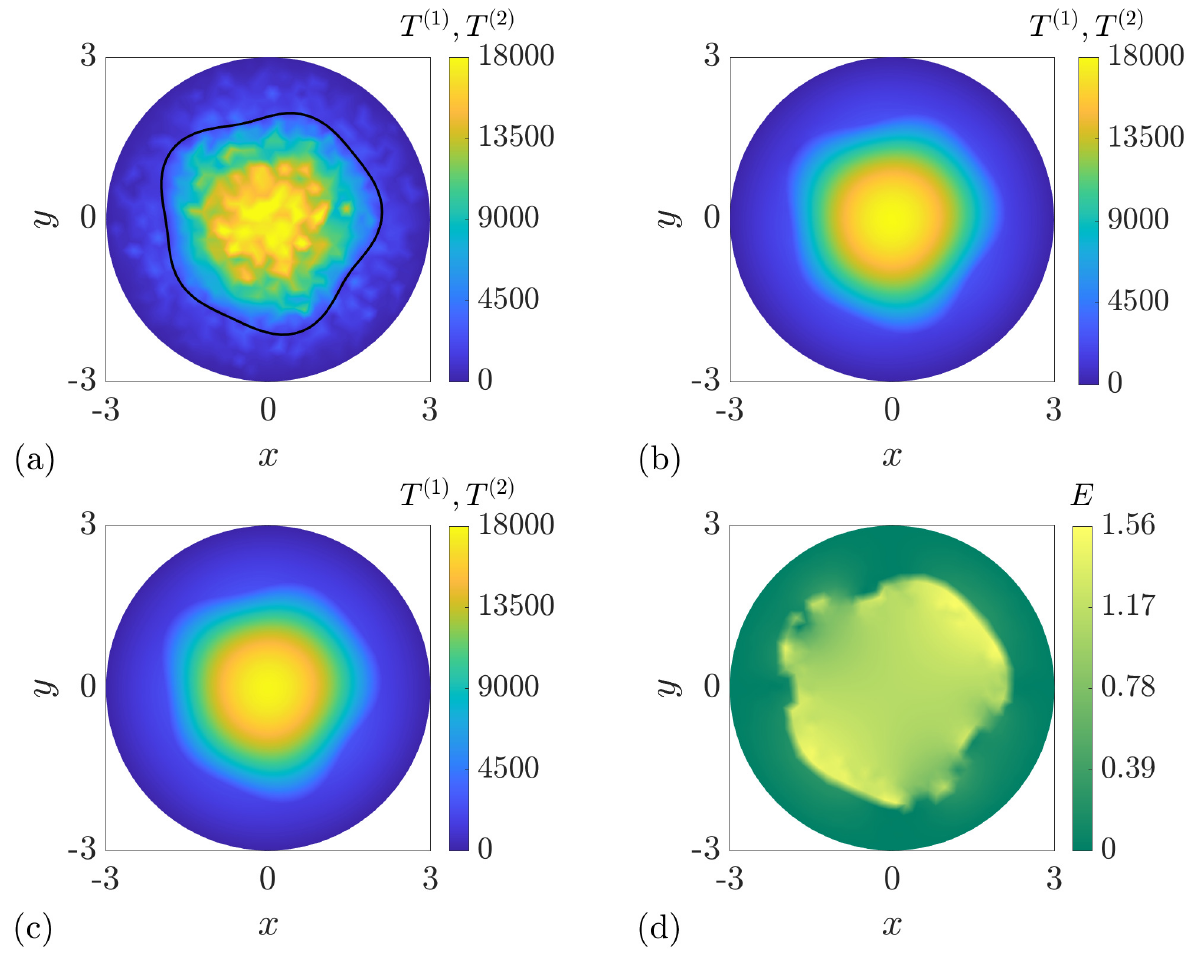}}
	\caption{MET on a compound disc with a perturbed interface. (a) stochastic simulations (b) perturbation solution defined in equations (\ref{eq:S_expansion})--(\ref{eq:T_expansion}) and (\ref{eq:T10})--(\ref{eq:T2_fourier}) (c) numerical solution \revision{(d) relative percentage difference between the perturbation and numerical solutions}. All results correspond to the parameter values $R_{1} = 2$, $R_{2} = 3$, $\varepsilon=0.05$, $g(\theta) = \sin(3\theta) + \cos(5\theta)$, $\delta = 0.05$, $\tau = 1$, $P_{1} = 1$, $P_{2} = 0.1$, $D_{1} = 6.25\times 10^{-4}$ and $D_{2} = 6.25\times 10^{-5}$. The results in (a) and (c) were produced using a mesh consisting of 928 nodes and 1759 triangular elements. The results in (b) were produced using the first three terms in the perturbation expansions (\ref{eq:S_expansion})--(\ref{eq:T_expansion}) and the first 25 terms in the Fourier series (\ref{eq:T1_fourier})--(\ref{eq:T2_fourier}). The black curve in (a) shows the interface.}
	\label{fig:compound_disc1}
\end{figure}

\begin{figure}[t]
	\centering
	\fbox{\includegraphics[width=0.8\textwidth]{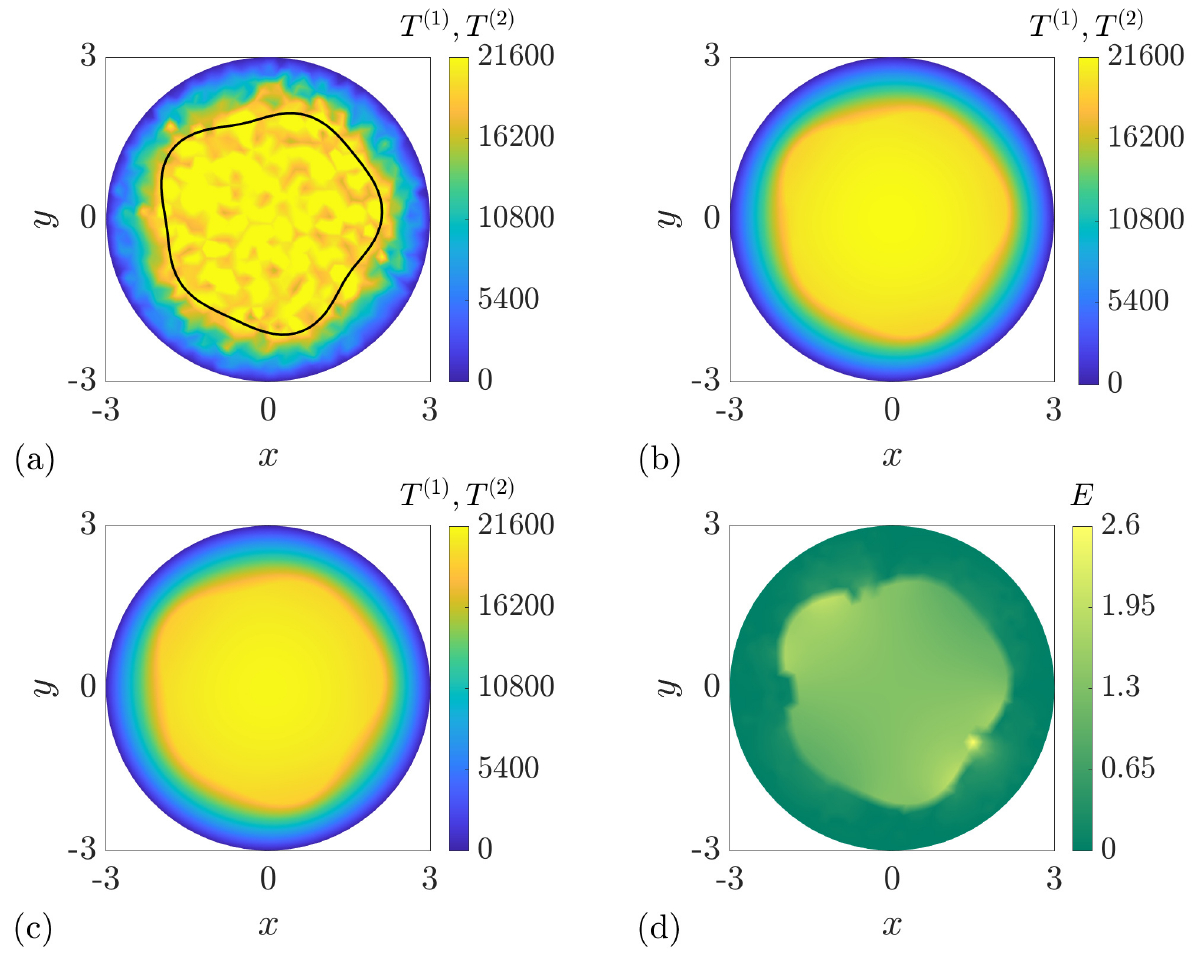}}
	\caption{MET on a compound disc with a perturbed interface. (a) stochastic simulations (b) perturbation solution defined in equations (\ref{eq:S_expansion})--(\ref{eq:T_expansion}) and (\ref{eq:T10})--(\ref{eq:T2_fourier}) (c) numerical solution \revision{(d) relative percentage difference between the perturbation and numerical solutions}. All results correspond to the parameter values $R_{1} = 2$, $R_{2} = 3$, $\varepsilon=0.05$, $g(\theta) = \sin(3\theta) + \cos(5\theta)$, $\delta = 0.05$, $\tau = 1$, $P_{1} = 0.1$, $P_{2} = 1$, $D_{1} = 6.25\times 10^{-5}$ and $D_{2} = 6.25\times 10^{-4}$. The results in (a) and (c) were produced using a mesh consisting of 928 nodes and 1759 triangular elements. The results in (b) were produced using the first three terms in the perturbation expansions (\ref{eq:S_expansion})--(\ref{eq:T_expansion}) and the first 25 terms in the Fourier series (\ref{eq:T1_fourier})--(\ref{eq:T2_fourier}). The black curve in (a) shows the interface.}
	\label{fig:compound_disc2}
\end{figure}

Results in Figures \ref{fig:compound_disc1} and \ref{fig:compound_disc2} compare the MET obtained from stochastic simulations to the MET obtained from the perturbation and numerical solutions for a perturbed compound disc with (i) low diffusivity inner region and high diffusivity outer region where $D_{1}/D_{2} = 0.1$ (Figure \ref{fig:compound_disc1}) and (ii) high diffusivity inner region and low diffusivity outer region where $D_{1}/D_{2} = 10$ (Figure \ref{fig:compound_disc2}). As with all previous results, the perturbation solution compares very well with the numerical solution and stochastic simulations.

\section{Conclusions and outlook} \label{Conc}
In this work we derive a suite of new exact solutions for the MET in a range of irregular geometries, including some situations where diffusion takes place in heterogeneous media.  Using appropriately averaged data from stochastic simulations and a numerical solution of the associated BVP, we provide evidence that the new exact solutions accurately describe the MET under a range of scenarios motivated by the biological images in Figure \ref{fig1}.  While the suite of results presented in Figures \ref{fig:annulus1}--\ref{fig:compound_disc2} cover many different cases, \revision{there are several other configurations that we have not explicitly considered such as a compound disc with a perturbed outer boundary (in addition to, or instead of, a perturbed interface) or a compound annulus with a perturbed interface (in addition to, or instead of, perturbed inner/outer boundaries).} All of these additional cases are conceptually straightforward extensions of the cases that we considered in this work. 

\revision{Extension of our perturbation solutions to other types of boundary conditions is also possible. For example, a perturbed annulus with a partially reactive inner boundary \cite{Grebenkov2010} and absorbing outer boundary can be accommodated by carrying out the following steps: (i) Replace the inner boundary condition (\ref{eq:ann2_T_inner}) with the Robin boundary condition $T(R_{1} + R_{1} \epsilon g_{1}(\theta), \theta) + \sigma \nabla T(R_{1} + R_{1} \epsilon g_{1}(\theta), \theta)\cdot\mathbf{n}(\theta) = 0$ where $\sigma > 0$. (ii) Replace (\ref{eq:ann2_T_inner_leading}) and (\ref{eq:ann2_T_inner_higher}) with the Robin boundary conditions obtained by introducing the perturbation expansion (\ref{eq:ann_T_expansion}) into the boundary condition from (i) and expanding in Taylor series centered at $r = R_{1}$. (iii) Redevelop the solutions for the leading order (\ref{eq:ann2_T0}) and higher order (\ref{eq:ann2_T_fourier})--(\ref{eq:ann2_T_fourier_Vn}) terms. For the compound disc, a similar repeat of the analysis allows imperfect contact \cite{Carr2018b,Moutal2019} at the interface to be accommodated. Here, to capture a discontinuity in MET at the interface caused by an infinitesimally-small interfacial barrier, the continuity condition (\ref{eq:pcd_ic1}) is replaced by $D_{1}\nabla T^{(1)}(\mathcal{R}_{1}(\theta),\theta)\cdot\mathbf{n}(\theta) = \sigma (T^{(1)}(\mathcal{R}_{1}(\theta),\theta) - T^{(2)}(\mathcal{R}_{1}(\theta),\theta))$ where $\sigma > 0$.}

In all cases in Figures \ref{fig:annulus1}--\ref{fig:compound_disc2} we consider a modest number of identically prepared stochastic simulations released from each location in the domain so that these results show some small fluctuations.  All software required to reproduce these results is provided on \href{https://github.com/ProfMJSimpson/Exit_time}{GitHub} where it is possible to vary the number of particles released at each location in the stochastic simulations.  Simply increasing the number of particles released at each location leads to smaller fluctuations in the MET, and these fluctuations eventually decrease to a point where it is impossible to visually distinguish between the averaged stochastic data, the numerical solution of the BVP and our perturbation solutions.  Instead of simply presenting a series of figures where each subfigure is visually indistinguishable, here we chose to work with a modest number of realizations of the stochastic simulations so that the fluctuations in the MET are visually clear.

\revision{In this work, along with our previous studies \cite{Simpson2021,Carr2022}, we focus on deriving closed-form expressions for the mean exit time on a range of two-dimensional domains, some of which include relatively simple forms of heterogeneity.  These approaches also apply to three-dimensional mean exit time problems where standard domains, such as spheres and compound spheres \cite{Carr2020}, can be treated in a similar way \cite{Piazza2019}, leading to closed-form expressions for the mean exit time in a range of perturbed three-dimensional domains, including some forms of heterogeneity.  One of the key challenges in these three-dimensional problems is implementing the finite volume numerical method to check the accuracy of the perturbation approximations, since meshing three-dimensional heterogeneous domains is significantly more challenging than working in two-dimensions.  To alleviate these issues, we anticipate reformulating the boundary value problem governing the mean exit time as a boundary integral problem will have significant advantages since we would then only need to discretise the surface of the domain, rather than the whole three-dimensional domain \cite{McCue1999,McCue2002}.  We hope to address this in future work.}

Overall, our results show that solving for the MET on irregular domains where the boundaries are small perturbations of regular shapes, such as an annulus and a compound disc, leads to very accurate approximations.  This approach can be used to study the MET in practical scenarios by digitizing irregular shapes and then approximating these shapes by choosing $g(\theta)$ (perturbed compound disc) and $g_1(\theta)$ and $g_2(\theta)$ (perturbed annulus) appropriately~\cite{Simpson2021}.  However, this approximation is not valid for domains that are not small perturbations of such regular geometries.  In the situation where we are interested in evaluating the MET on a domain that is not a small perturbation of a regular shape we must rely on repeated stochastic simulation data and numerical solutions of the associated BVP until new approximations for these conditions are developed.  We hope to develop such approximations in future research.\\

\noindent
\textit{Acknowledgements:} This work is supported by the Australian Research Council (DP200100177). We also acknowledge funding from Queensland University of Technology which provided DJV with a 2020-2021 Vacation Research Experience Scheme (VRES) stipend. \revision{We thank the two anonymous referees for their helpful comments.}

\appendix
\counterwithin{figure}{section}
\numberwithin{equation}{section}

\section{Numerical solutions}
\label{app:finite_volume_solution}
Throughout this work analytical solutions of Equation (\ref{eq:GovEq}) are visually compared to and benchmarked against numerical solutions. These numerical solutions are obtained as discussed in our previous work \cite{Simpson2021} by using a finite volume discretisation of Equation (\ref{eq:GovEq}) on an unstructured triangular meshing of $\Omega$. The finite volume method is implemented using a vertex-centered strategy where nodes are located at the vertices in the mesh and control volumes are constructed around each node by connecting the centroid of each triangular element to the midpoint of its edges~\cite{Carr2016}. Meshes are generated using the mesh generation software GMSH~\cite{Geuzaine09} and linear interpolation is used to approximate gradients in each element. Assembling the finite volume equations yields a sparse linear system whose solution provides point estimates of the MET, $T(\mathbf{x})$, at each node in the finite volume mesh. These point estimates are then visualised across $\Omega$ using MATLAB's \texttt{trisurf} function. A \matlab\ implementation of our numerical algorithm is available on \href{https://github.com/ProfMJSimpson/Exit_time}{GitHub}. 

\revision{\section{Error calculation}
\label{app:error}
To quantify the discrepancy between the perturbation and numerical solutions, we follow our previous work \cite{Simpson2021} and calculate point values of the relative percentage error, $E(\mathbf{x}) = 100|T_{\mathrm{p}}(\mathbf{x})-T_{\mathrm{n}}(\mathbf{x})|/\max_{\mathbf{x}\in\Omega}T_{\mathrm{n}}(\mathbf{x})$, at each node in the finite volume mesh. Note that subscripts are introduced here to distinguish between the perturbation ($\mathrm{p}$) and numerical ($\mathrm{n}$) solutions. These point values of $E(\mathbf{x})$ are then visualised across $\Omega$ using MATLAB's \texttt{trisurf} function.}

\section{Stochastic simulations}
\label{app:stochastic_model}
Throughout this work  analytical solutions of Equation (\ref{eq:GovEq}) are visually compared to estimates of MET from repeated stochastic simulations. These stochastic simulations estimate $T(\mathbf{x})$ using $N \gg 1$ identically-prepared random walk realisations for each position $\mathbf{x}$. Each realisation involves releasing a particle at position $\mathbf{x}$ and recording the elapsed time until the particle passes through a designated absorbing boundary. The random walk algorithm involves releasing a particle at $\mathbf{p}(0)=\mathbf{x}$, at time $t = 0$, and proceeds by simulating a series of constant spatial steps of length $\delta > 0$ across a series of constant time steps of duration $\tau > 0$. For a particle located at position $\mathbf{p}(t)\in\Omega$ at time $t$ the possible outcomes during the time step from $t$ to $t + \tau$ follow our previous work \cite{Carr2018,Carr2020}. For the annulus and perturbed annulus the particle either: (i) moves from its current position, $\mathbf{p}(t)$, to a new position $\mathbf{p}(t+\tau) = \mathbf{p}(t) + \delta(\cos(\theta),\sin(\theta))$ where $\theta\sim\mathcal{U}[0,2\pi]$ with probability $P \in [0,1]$; or (ii) remains at the current position, $\mathbf{p}(t+\tau) = \mathbf{p}(t)$, with probability $1-P$. For the compound disc and perturbed compound disc, the particle either: (i) moves to one of $n$ positions: $\mathbf{p}(t+\tau) = \mathbf{p}(t) + \delta(\cos(\theta_{k}),\sin(\theta_{k}))$ ($k=1,\hdots,n$) where $\theta_{k} = 2\pi (k-1)/n$ with probability $\mathcal{P}_{k}/n$; or (ii) remains at the current position, $\mathbf{p}(t)$, with probability $1-\sum_{k=1}^{n}\mathcal{P}_{k}/n$, where we take $n=24$ \cite{Carr2020}. Here, $\mathcal{P}_{k} = P_{1}$ if $\mathbf{p}(t) + (\delta/2)(\cos(\theta_{k}),\sin(\theta_{k}))$ is located in the inner region ($0 < r < R_{1}$ for the compound disc and $0 < r < \mathcal{R}_{1}(\theta)$ for the perturbed compound disc) and $\mathcal{P}_{k} = P_{2}$ if $\mathbf{p}(t) + (\delta/2)(\cos(\theta_{k}),\sin(\theta_{k}))$ is located in the outer region ($R_{1} < r < R_{2}$ for the compound disc and $\mathcal{R}_{1}(\theta) < r < R_{2}$ for the perturbed compound disc). For all geometries, if the potential step requires the particle to pass through a designated reflecting boundary the step is aborted.

In summary, for a given starting position $\mathbf{x}$, this gives us $N$ estimates of the exit time from which we calculate the mean, $T(\mathbf{x}) = (1/N) \sum_{i=1}^{N} t_{i}$, where $t_i$ is the exit time from the $i$th identically-prepared stochastic realisation. Repeating this process for every node in the finite volume mesh yielding point estimates of the MET, $T(\mathbf{x})$, at each node. These point estimates are then visualised across $\Omega$ using MATLAB's \texttt{trisurf} function with all results in this work corresponding to $N = 50$ simulations.   This means that for a mesh with $M$ nodes, we perform $N \times M$ random walk simulations, for example, for the results in Figures \ref{fig:annulus1}--\ref{fig:annulus3} with $M=986$ nodes and $N=50$ simulations per node, our estimates of $T(\mathbf{x})$ are obtained using $N \times M = 28650$ stochastic simulations. A \matlab\ implementation of our simulation algorithm is available on \href{https://github.com/ProfMJSimpson/Exit_time}{GitHub}, and these algorithms can be used to estimate $T(\mathbf{x})$ with different choices of $N$, where it becomes clear that the fluctuations in $T(\mathbf{x})$ decay as $N$ increases.

\end{document}